%% file: Draft_vFinal.tex
\documentclass[final]{IEEEtran}
\usepackage{amsthm,amssymb,graphicx,multirow,amsmath,color,amsfonts}
\usepackage[update,prepend]{epstopdf}
\usepackage{cite}
\usepackage[latin1]{inputenc}
\usepackage{pdfpages}
\usepackage{tabulary}
\usepackage{multirow}
\usepackage{mathtools}
\usepackage{caption}
\usepackage{subcaption}

\usepackage{duckuments}

\usepackage{algorithm2e}

\SetKw{algobreak}{break}
\usepackage{mathtools, cuted}
\usepackage{lipsum}

\include{notation}

\allowdisplaybreaks 
\begin{document}
\graphicspath{{./Figures/}}
\title{
Charging Techniques for UAV-assisted Data Collection: Is Laser Power Beaming the Answer? }

\author{Mohamed-Amine~Lahmeri,~\IEEEmembership{Student Member,~IEEE,}
	Mustafa~A. Kishk,~\IEEEmembership{Member,~IEEE,}
	and~ Mohamed-Slim~Alouini,~\IEEEmembership{Fellow,~IEEE}
	\thanks{Mohamed-Amine Lahmeri is a Ph.D. student at the Institute for Digital Communications (IDC), Friedrich-Alexander-Universität Erlangen-Nürnberg, Germany, and carried this work while being at the Communication Theory Lab (CTL), King Abdullah University of Science and Technology (KAUST); Mustafa A.Kishk and Mohamed-Slim Alouini are with the Computer, Electrical, and Mathematical Science and Engineering (CEMSE) Division, KAUST, Thuwal 23955, Saudi Arabia (e-mail: amine.lahmeri@fau.de; mustafa.kishk@kaust.edu.sa;
		slim.alouini@kaust.edu.sa).}
}

\maketitle
\begin{abstract}
	As Covid-19 has increased the need for connectivity around the world, researchers are targeting new technologies that could improve coverage and connect the unconnected in order to make progress toward the United Nations Sustainable Development Goals. In this context, drones are seen as one of the key features of 6G wireless networks that could extend the coverage of previous wireless network generations. That said, limited on-board energy seems to be the main drawback that hinders the use of drones for wireless coverage. Therefore, different wireless and wired charging techniques, such as laser beaming, charging stations, and tether stations are proposed. In this paper, we analyze and compare these different charging techniques by performing extensive simulations for the scenario of drone-assisted data collection from ground-based Internet of Things (IoT) devices. We analyze the strengths and weaknesses of each charging technique, and finally show that laser-powered drones strongly compete with, and outperform in some scenarios other charging techniques. 
\end{abstract}

\vspace{-2mm}
\section{Introduction} \label{sec:intro}

With the roadmap toward the integration of drones into today's applications being formulated, the value of the global drone market, estimated at 27 USD billion, is expected to reach  58 USD billion by 2026, showing how fast the drone industry is evolving. As a result, drones have started to open up new opportunities for several applications such as mapping, inspection, emergency, agriculture, and recently wireless communications~\cite{1,11}. Multiple striking features of drones have helped them take the spotlight. These include their flexibility, ease of deployment, and ability to establish a line of sight (LOS) with ground targets.  These characteristics have also encouraged the deployment of drones for wireless communication purposes. Thus, several research studies have focused on the use of Unmanned Aerial Vehicles  (UAVs) as airborne base stations and relays to extend coverage~\cite{2}. Other studies have focused on drone-assisted data collection for Internet of Things (IoT) purposes, especially with the widespread adoption of connected devices around the world~\cite{3}.\par
While the integration of drones into existing wireless networks looks promising, powering drones is becoming a serious problem. Indeed, most commercially available drones have difficulty staying in the air for more than 30 minutes. That being said, in order to use drones as base stations or aerial relays, their operating time needs to be significantly increased. In this context, optimization techniques targeting trajectory planning, battery, or resource allocation~\cite{6,9,10} could increase the mission time of drones, but they are still limited.  Artificial intelligence techniques have also been used to optimize the trajectory and resources of drones when operating as base stations or air relays, such as machine learning, deep learning, reinforcement learning, and federated learning~\cite{11}. Although several interesting results have been provided, they are still considered limited. Added to this are the various challenges in using artificial intelligence onboard drones.  For this reason, researchers are investigating different wireless and wired charging techniques for drones to help them in long-duration missions. One potential solution is to equip drones with solar panels to harvest energy from daylight~\cite{15}. However, in addition to its limited effectiveness, this technique can only work during the day. This makes it unsuitable for wireless communication missions.  Battery swapping is another charging technique that involves deploying multiple charging stations for drones, where their battery is replaced when it reaches critical values. This technique is usually assisted by a human and requires additional time to perform the battery swap. It also requires multiple batteries for a single drone, which increases the cost of this technique. The tethered drone represents another way to provide indefinite power to drones by connecting the drone to a ground station via a cable~\cite{13}. The physical link between the ground station and the drone not only transfers power but also provides a reliable backhaul to the drone that can be used to offload its data. However, tethering the drone limits its mobility and can affect its performance, especially for wireless coverage improvement missions. This also raises the question of what type of tether to use and the complexity of manufacturing lightweight, efficient cables. In addition, drone swapping is another alternative to assist in long-duration missions, however, this technique requires multiple drones for a mission, which increases the cost of the solution. It is also important to mention that coordination is of paramount importance in such scenarios.\par 
	Wireless Power Transfer (WPT) is a novel solution for providing indefinite power to drones without having to visit charging stations or connect to a physical cable. This could be achieved by using laser beams to transmit high power to the drone over large distances\cite{6,10}.  In this context, some companies have already proven the feasibility of such a system by powering a drone for more than 48 hours. For example, PowerLight Technologies has provided a working prototype of laser-powered drones~\cite{12}.  In addition, several publications have targeted the use of laser-powered drones for wireless communication scenarios. The authors of~\cite{4} recently investigated different commercial photovoltaic materials used to harvest energy on the laser-powered drone. In~\cite{5} the authors highlighted three different charging techniques, namely drone swapping, battery swapping, and laser charging. The weaknesses and advantages of each charging technique were highlighted. In this paper, we highlight the large-scale deployment of ground-based laser beam directors (LBDs) as a promising alternative for wireless communication-based missions. In addition, we consider a drone-assisted IoT data collection scenario. We show that the future use of laser-powered drones could outperform other traditional charging techniques. To this end, we compare the performance of laser-powered drones to tethered and untethered drones. We also highlight the weaknesses and strengths of these charging techniques and discuss challenges and open problems. 
\section{Laser beaming technologies for UAVs}
\begin{figure*}
	\centering
	\includegraphics[width=\textwidth]{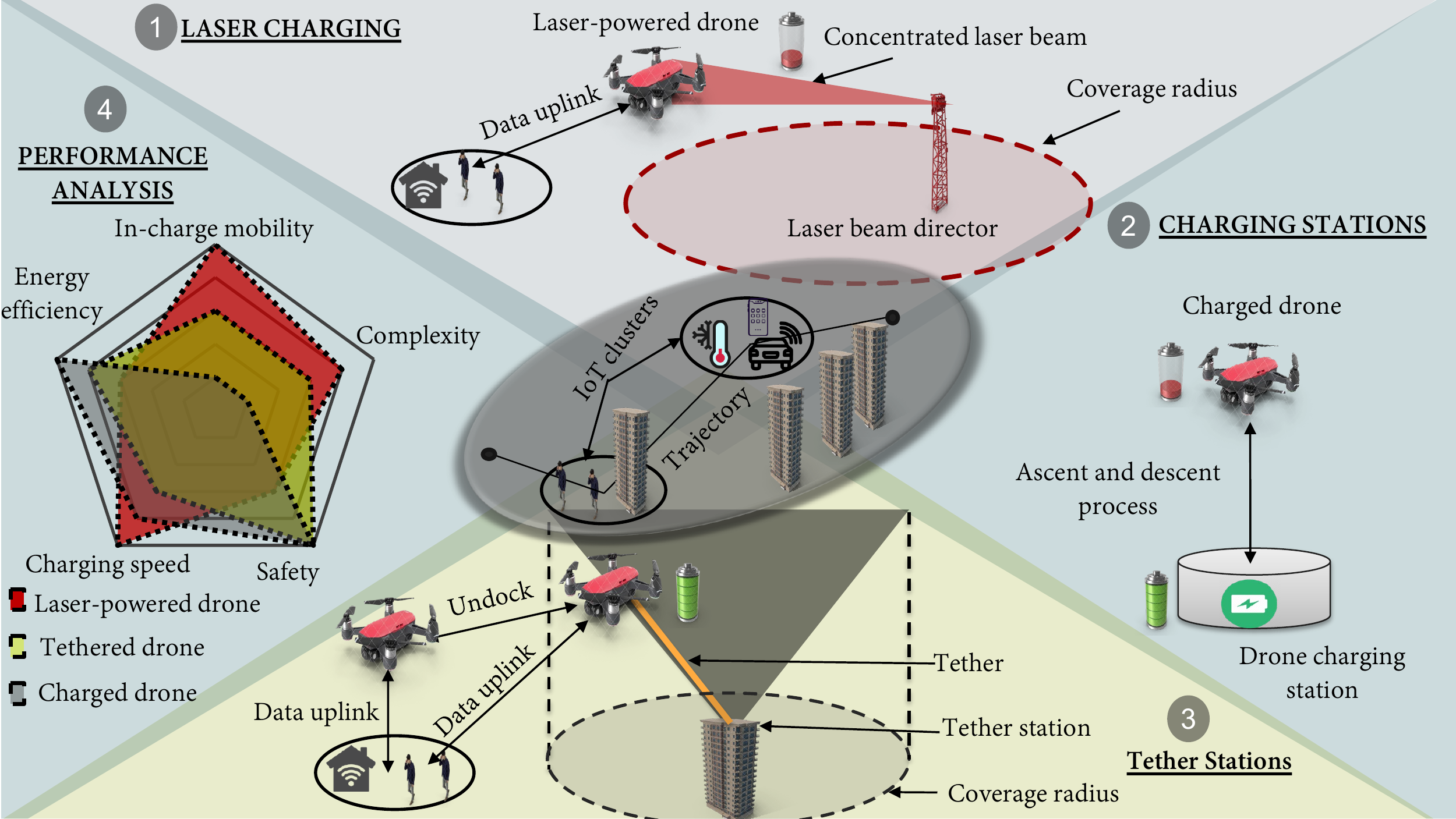}
	\caption{An overview on the different drone charging techniques. }
	\label{fig:0}
	\vspace{-3mm}
\end{figure*}
Historically speaking, after its invention in 1960, laser light has been used in a plethora of applications, not limited to communication but also in medicine, nuclear fusion, radar systems, to name a few~\cite{14}. Inspired by the ideas of Nikolas Tesla, laser light has also been used to transfer energy wirelessly over great distances. Recently, with the remarkable evolution of the drone market and the rapid advent of their technologies, scientists are looking to power flying platforms with laser beams for indefinite periods of time. As mentioned earlier, a working prototype has already been implemented.  Generally speaking, the mechanism described consists of a ground-based optical system equipped with an array of high-power laser sources, a set of mirrors, a cooling system, a safety system, and a tracking system to direct the laser beam to the UAV~\cite{12}. As for the laser source, it usually consists of a high-powered laser array, with power reaching several kilowatts. In~\cite{12}, a diode-pumped fiber laser with an emission power of 4 kW is mentioned. As part of their multi-year PATROL project, and in cooperation with the U.S. Naval Research Laboratory, PowerLight Technologies made a remarkable demonstration of indoor power beaming as part of the second phase of the project. The power of the laser source used was 2 kW and provided more than 400 W of direct current at a distance of 325 m. Recently, the third phase of the project was launched, where the energy will be received by a smaller photovoltaic cell, light enough to be equipped onboard. The authors of\cite{4} used an 850 nm laser source with an emission power of 600 W.     Since high power is used to deliver energy at a long range, cooling systems must be deployed to adjust the temperature of the whole system to normal values. Regarding safety, it should be mentioned that the use of laser beams outdoors can be dangerous if directed at human eyes. Although the altitude of the drone greatly reduces the risk of hitting human eyes, the safety system increases the overall safety level of the system by checking for the presence of a LOS between the ground module and the drone. The laser source is turned off as soon as something comes near the laser beam. On the other hand, the drone is equipped with a photovoltaic receiver, sensitive to the wavelength of the emitted laser, and able to efficiently harvest the energy of the laser beam and convert it into an electric current. 
\section{A use case scenario:}

In is paper, we consider a drone-assisted uplink scenario from uniformly distributed ground-based IoT clusters. As described in Fig.~\ref{fig:0}, each cluster consists of multiple connected devices with different amounts of data. The drone's mission consists of hovering over these clusters, collecting data, and offloading it to a server.  To prolong the UAV's mission time, we consider deploying LBDs on the ground. The LBDs provide energy to the UAV based on distance as explained in~\cite{7}. Therefore, the shorter the distance, the faster the charging time. We aim to compare the performance of the laser-powered drone to other types of drones, which we present in the next section. To this end, we perform extensive simulations and compare the average performances in the different case scenarios.
\subsection{Benchmarks:} 
We compare the laser-powered drone to mainly 3 types of drones described as follows:
\begin{itemize}
\item Non-charged drone: This corresponds to the baseline case of a regular rotary-wing UAV with an ordinary commercial battery onboard. The mission time of this drone is directly linked to the capacity of its battery as there are no charging capabilities. We want to compare regular drones to drones with charging capabilities to see if this type of drone is suitable for wireless communication-based missions. 
\item Charged drone: We consider randomly deploying charging stations on the ground such that the UAV can visit them to recharge its battery. Hence, if the UAV's battery is at critical levels, the UAV visits its nearest charging stations and initiates the recharging process. This process includes a descent phase where the drone lowers its altitude until it reaches the charging platform, recharges its battery, and finally an ascent process to return to an operational altitude.  We consider neither battery swapping nor UAV swapping since we aim at performing fair comparisons between the different UAV charging techniques. In the case where the drone cannot reach the nearest charging distances, especially when the density of charging stations is low, it performs an emergency landing and terminates its missions after consuming its battery.
\item Tethered drone:
As the name implies, a tethered drone is a drone that is connected by a physical cable or tether to a power station on the ground. This physical connection between the drone and the ground power supply allows the drone to keep flying for long period and makes it suitable for wireless communication purposes. However, this comes with the cost of limiting the drone mobility, and hence the drone can only serve users that are within its coverage range. In this work, we propose equipping the tethered drone with the capability of docking and undocking to a tether, thus relying on its battery while moving from a tether station to another.  We believe this will allow the drone to not only serve users within the range of the tether but also users that are located outside that range, relying solely on its battery. Docking and undocking procedures can be human-assisted and require lowering the drone's altitude until it reaches the tethering station, then returning it to its operational altitude once docked. As in the previous cases, we assume that the docking stations are randomly deployed on the ground. 
\end{itemize}
\subsection{Performance Metrics:}
To make reasonable comparisons, we rely on Monte Carlo simulations by averaging the drone's performance under different charging techniques. In each realization, we consider a random spatial distribution of the ground power supply stations, however, we keep the same locations for all the different scenarios in order to make fair comparisons later on. To this end, we use different performance metrics to quantify the quality of service provided by the drone. For example, we rely on data harvesting efficiency as the primary performance metric that reflects the quality of service provided by the drone. Data harvesting efficiency is defined as the number of fully served IoT clusters relative to the total number of available IoT clusters in the simulation area. Thus, a percentage of 100\% refers to the case where the drone successfully collected all available data in the area. Another performance indicator is the required number of power stations to achieve a certain level of data harvesting efficiency. In addition, we also rely on another energy-related performance metric, namely the average distance traveled by the drone. In fact, it is important to track the total distance covered by the drone to serve a given number of IoT ground clusters. This can also give us an idea of the energy consumed during the mission, as the uplink power is negligible compared to the propulsion power of the drone. 
\section{Numerical Results}
\subsection{Simulation Setting}
We consider a UAV-assisted data collection from ground IoT clusters on the ground. We assume that the users have the same priority level, so the drone prioritizes the nearest unserved cluster and starts the uplink as soon as the users are in its coverage area, where its radius is denoted by $r_{\rm cov}$. We consider a commercial rotary-wing drone equipped with a 4 Series(4S) Lithium-Polymer (LiPo) battery. The drone is hovering at an operational altitude of 100m and with an optimal velocity that minimizes its energy consumption as in~\cite{8}. The UAV gets to the ground either for charging or following an emergency landing. In addition, the laser-powered drone and the tethered drone can harvest energy while collecting data from users and vice versa.  We consider a total of 18 clusters deployed on the ground, however, the charging stations are limited to 6 units in the considered area. Both the charging station and the clusters are randomly scattered on the ground. To perform trajectory planning we discretize the maximum mission allowed time into small intervals, denoted by $\Delta T$, where the drone at each time slot decides what direction to choose based on its distance to the nodes and the state of its battery. We summarize all the numerical values into Table~\ref{Table1}. These are the values used in the figures unless otherwise stated. Other values related to the laser link, or to the UAV dynamics can be found in~\cite{7,8}.
\subsection{Performance results}
\begin{figure}
	\centering
	\includegraphics[width=0.95\columnwidth]{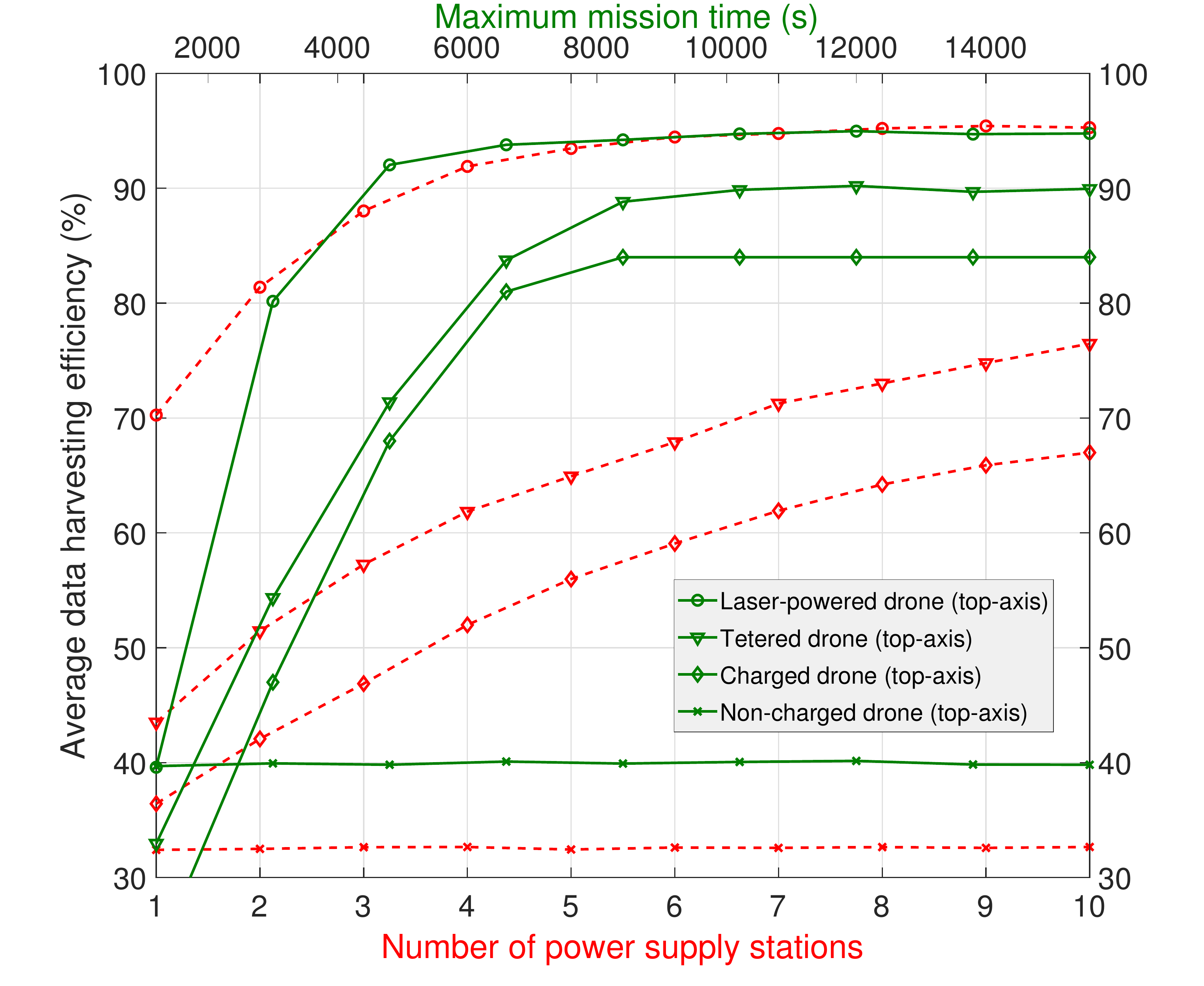}
	\caption{Data harvesting efficiency.}
	\label{fig:1}
	\vspace{-3mm}
\end{figure}
Fig.~\ref{fig:1} shows the average data collection efficiency as a function of the number of ground power supply stations and the maximum duration of the IoT data collection mission. The solid green curves correspond to the plot against the maximum mission duration represented by the upper x-axis, while the red dashed lines correspond to the plot against the number of ground power supply stations represented by the lower x-axis.   As shown with the green solid lines, when the drone is not equipped with any charging capability, the data harvesting efficiency is not affected by the maximum mission duration, as expected. This is because the drone serves the nearest IoT clusters as long as its battery allows that, and then initiates an emergency landing procedure. As a result, the average data harvesting efficiency is about 40\% in this scenario, which shows how limited commercial drones are, especially for wireless-based missions. This is not the case for drones equipped with charging capabilities. As shown in the figure, laser-powered drones outperform all other types of drones by achieving an average data collection efficiency of 95\%. In addition, drones equipped with ground charging stations are slightly more efficient than tethered drones, but the efficiency can still be considered comparable. This slight improvement is due to the fact that the charging procedure for tethered drones is more complicated than for charged drones.  This is because the charging procedure for tethered drones includes a docking phase with an ascent and descent process and then a second undocking phase with similar steps. In addition, the mobility of tethered drones is limited when they are attached to a docking station. When it comes to the effect of the ground charging stations plotted with red dashed lines, we observe an improvement in drone performance as the number of ground charging stations increases, except for regular uncharged drones. We also notice that the performance stabilizes from a certain number of charging stations that we consider optimal. For example, the laser-powered drone, which is the best in terms of uplink performance, achieves almost constant performance when there are 6 LBDs in the total area considered. Charged drones are still better than tethered drones, but both require more ground charging stations than laser-powered drones. This is mainly because the coverage radius of LBDs is remarkably wide, so fewer stations are needed to provide optimal spatial coverage.\par
\begin{figure}
	\centering
	\includegraphics[width=\columnwidth]{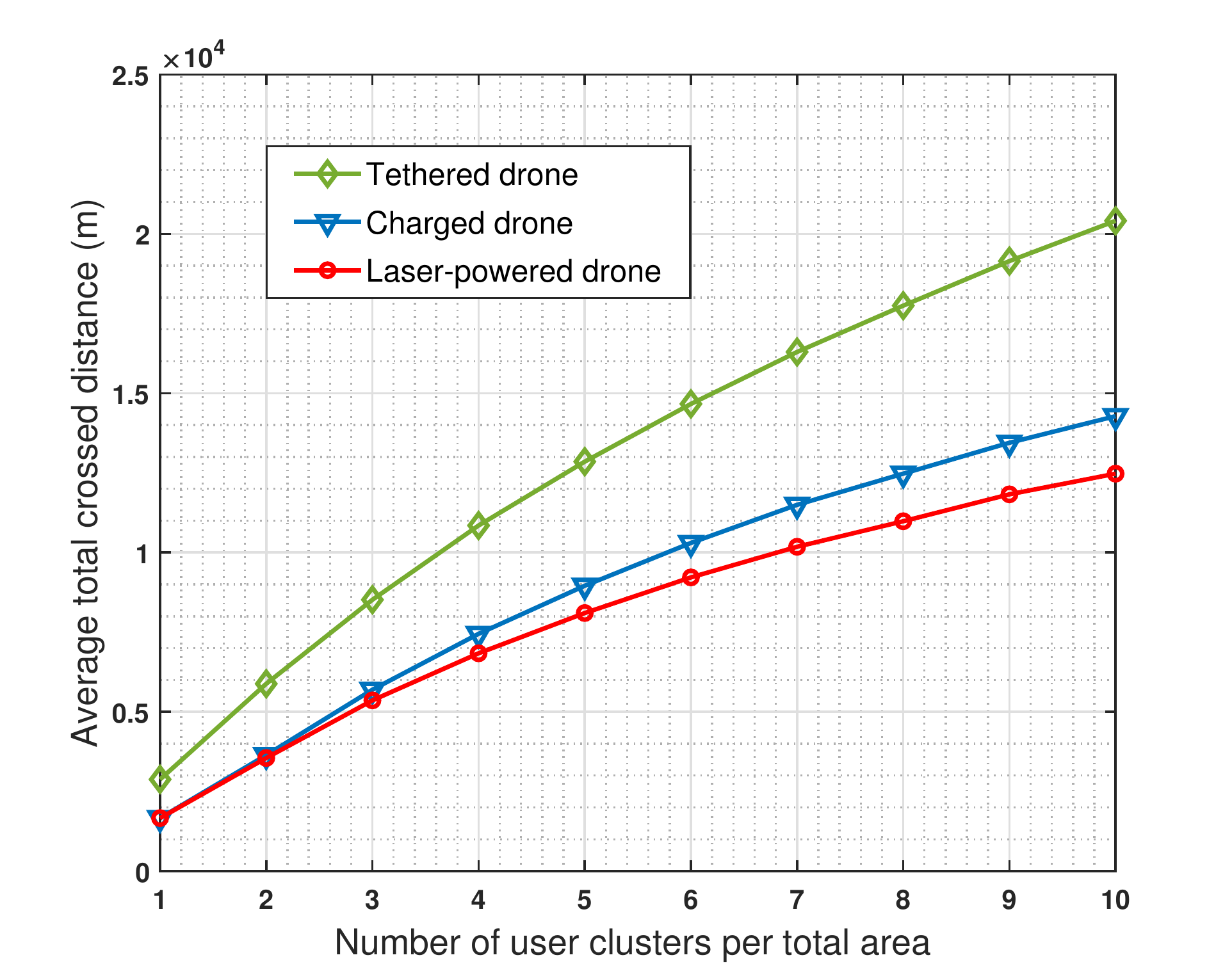}
	\caption{Drone movement efficiency.}
	\label{fig:2}
	\vspace{-3mm}
\end{figure}

Fig.~\ref{fig:2} plots the average movement efficiency of the drone against the total number of clusters in the area. In other words, it tracks the average distance the drone has to travel to serve a certain number of IoT clusters on the ground. This number could also give us an idea of the propulsion energy consumed by the drone. It is clear that laser-powered drones are so far the most efficient type of drone in terms of average distance traveled. This is due to the fact that this type of drone has the ability to recharge its batteries wirelessly, without the need to visit LBD locations. In addition, it is possible to serve clusters while harvesting energy. On the other hand, charged drones are less efficient, which can be explained by the fact that this type of drone not only has to visit the exact locations of the charging stations but also has to lower its altitude and return to the operational altitude, which is counted in the total distance traveled. That being said, tethered drones perform poorly in terms of distance traveled, as shown by the solid green line in Fig.2. This performance is expected because this type of drone must first move to a tether station in order to serve users, which requires a docking and undocking process. These operations are costly in terms of distance traveled and time, as shown in the figure.\par

\begin{figure*}
	\centering
	\includegraphics[width=\textwidth]{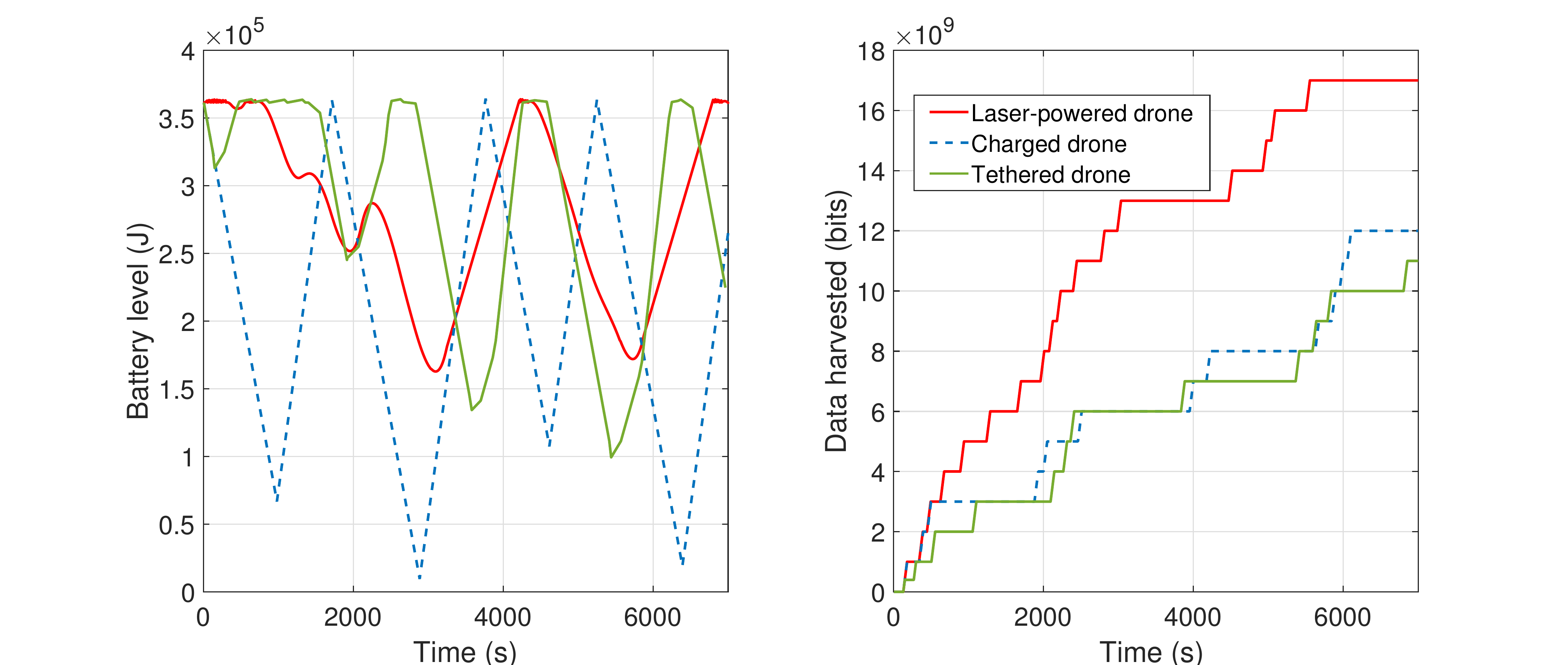}
	\caption{The dynamic profile of the drone battery and harvested data.}
	\label{fig:4}
	\vspace{-3mm}
\end{figure*}
Unlike the previously commented figures where the average performance was revealed, Fig.~\ref{fig:4} captures the dynamic profile of the different types of drones by tracking their battery levels and the amount of data embedded in the drone at each time interval $\Delta T$.  On the left side of the figure, the battery profile shows the charge cycles for the different drone types. Charged drones have full charge cycles, while laser-powered drones benefit from continuous charging while hovering in the LBD coverage area. The charging speed also depends on the distance between the drone and the LBD, as shown by the solid red line.  The horizontal lines in the tethered drone charge cycles correspond to the docking and undocking procedures. On the right side of the figure, we can see the real-time performance of the drones in terms of data collected in a limited time window.  Again, we can notice that the laser-powered drones outperform the other types of drones by harvesting more than 17 Gbits in a time window of about an hour and a half. In a time window of 7000 seconds, the charged drone managed to harvest 12 Mbits while the tethered drone harvested 11 Mbits of data. 
\begin{table}[]\caption{Numercial Values}
	\label{Table1}
	\centering
	\resizebox{0.5\textwidth}{!}{%
		\begin{tabular}{|c|c|c|c|}
			\hline
			Parameter               & Numerical value & Parameter               & Numerical value \\ \hline
			\multicolumn{2}{|c|}{Simulation}          & Battery capacity        & $6700$ mAh        \\ \hline
			Area                    & $25 $   km$^2$   & Battery nominal voltage & $14.8 V$ V          \\ \hline
			Trials                  & $10^4$          & Uplink data rate        & $20$ Mbps        \\ \hline
			Time window             & $9000 $  s      & Operational altitude    & $100$ m            \\ \hline
			$\Delta T$ & $1s$            & Velocity                & 6.2 m/s         \\ \hline
			Number of clusters      & $18$              & \multicolumn{2}{c|}{Charging stations}    \\ \hline
			\multicolumn{2}{|c|}{Drone}               & Number/area             & $6$               \\ \hline
			Weight                  & $1500$ g          & Docking time            & $120 $  s          \\ \hline
			Battery weight          & $500$ g           & Laser source power      & $2 $ kW           \\ \hline
		\end{tabular}
	}
\end{table}

\section{Challenges and discussions:}
In this paper, we highlighted what a large-scale deployment of laser-powered drones can bring to wireless communication-based missions. Although in terms of time and service performance, laser-powered drones outperform other types of drones, several challenges can hamper their concrete deployment in the future. In what follows, we delve into the details of some of these obstacles.
\begin{itemize}
\item Safety issues: 
Although most companies in this field claim that this technology is safe, it is still not clear that deploying these laser beam directors in crowded areas is completely safe. This could also pose an environmental problem, especially considering that not only humans are at risk, but all flying species as well. The question here is how accurate is the safety system and how reliable is it?  The latter question might open new research directions evaluating and improving the reliability of such systems. 
\item Energy efficiency:
When it comes to the energy aspect of laser-powered drones, we believe that all the remarkable performance of this type of drone has an energy cost. Indeed, when transmitting energy in free space, it is difficult if not impossible not to waste energy halfway. This is why LBDs use high power at the laser source. Thus, it is not clear that the performance gain can cover and justify the additional energy costs. At this level, the one positive point we can mention is that industrial companies are studying how to increase the efficiency of energy transfer by minimizing energy loss as much as possible. This is for example an integral part of the PATROL project of PowerLight Technologies. Thus, future studies should target further improving lightweight photodiode receivers' efficiency in harvesting laser energy. 
\item Technological cost and complexity:
Another challenge facing the large-scale deployment of LBDs is the cost and complexity of manufacturing the entire system used to provide power. The technological cost of such a complex laser source system can raise multiple concerns about the scalability of the system. While we believe this technology is still in the research and development phase, we believe that, as with any other technology product, the cost will decrease as the technology matures. So it will likely be some time before we are ready to deploy laser-powered drones on a large scale and for civilian purposes. 
\item Vulnerability to weather conditions:
Weather conditions impose another concrete challenge for laser-powered drones. In this context, authors in~\cite{7} investigated the effect of atmospheric turbulence on the laser beam. However, laser light is not only affected by atmospheric turbulence but also probably by weather conditions in general. For instance, the existence of smog, fog, snow, rain, and dust could affect the beam divergence and power at the drone. Therefore, future research might target the effect of weather conditions on drone laser charging. Moreover, weather conditions might not only affect the laser beam but also the tracking system. In other words, the real-time tracking of the drone movement might also be remarkably affected.
In a nutshell, the question of whether laser-powered drones are worth deploying in future generations of wireless communication remains complex. Quantifying the tradeoff between wireless efficiency and energy loss could go a long way toward answering our questions. That said, we believe that laser-powered drones are the best in terms of flexibility, charging speed, and quality of service, as shown in Fig.~\ref{fig:0}. However, this type of drone, as explained earlier, could lose points in terms of energy efficiency and overall system safety. In this context, tethered drones and rechargeable drones are believed to be much more energy-efficient. Moreover, system complexity is another issue for laser-powered and tethered drones compared to rechargeable drones that benefit from a simpler charging operation. At this point, we should bring up that all the problems related to laser power can be mitigated with the advent of technology, as the energy loss could be minimized in the coming years, and the same goes for the complexity of the system. 
\end{itemize} 

\section{Conclusion}

In this paper, we motivate a large-scale deployment of laser-powered drones by highlighting their performance compared to other charging techniques. Based on extensive Monte Carlo simulations, we show that laser-powered drones outperform tethered and recharged drones in terms of quality of service, distance traveled, and service time. Time-varying graphs showing the energy profile of the batteries and the amount of data collected were also provided.  With that said, we also discussed the challenges that might be faced with what we envision for laser-powered drones. Based on these challenges, we highlighted possible future research directions and open problems in this area. In summary, we believe that laser-powered drones will offer exciting new opportunities for wireless communications. This will be possible by taking advantage of the fact that the laser beam is silent, transparent, and precise, which will allow a smooth transfer of energy and information to flying drones.

\bibliographystyle{IEEEtran}
\bibliography{hokie-HD}

\vspace{2cm} 

\begin{IEEEbiographynophoto}
	{Mohamed Amine Lahmeri}
	[S'21]  was born in Tunis, Tunisia. He received the
	National Engineering Diploma degree from Ecole
	Polytechnique de Tunisie, Tunis, Tunisia, and M.Sc. from King Abdullah University of Science
	and Technology, Thuwal, Saudi Arabia.  He is currently pursuing
	a Ph.D. degree with the Institute for Digital Communications (IDC), Friedrich-Alexander-Universität Erlangen-Nürnberg, Germany. His current research interests include UAV-enabled communication systems, laser-powered drones, and machine learning for
	wireless communication.
\end{IEEEbiographynophoto}

\vspace{-1.2cm}

\begin{IEEEbiographynophoto}
	{Mustafa A. Kishk} [S'16, M'18] received the
	B.Sc. and M.Sc. degrees from Cairo University,
	Giza, Egypt, in 2013 and 2015, respectively, and
	the Ph.D. degree from Virginia Tech, Blacksburg,
	VA, USA, in 2018. He is currently a Postdoctoral
	Research Fellow with the Communication Theory
	Lab, King Abdullah University of Science and
	Technology, Thuwal, Saudi Arabia. His current research interests include stochastic geometry, energy harvesting wireless networks, UAVenabled communication systems, and satellite
	communications.
\end{IEEEbiographynophoto}

\vspace{-1.2cm}

\begin{IEEEbiographynophoto}
	{Mohamed-Slim Alouini} [S'94, M'98, SM'03, F'09] was
	born in Tunis, Tunisia. He received the Ph.D.
	degree in electrical engineering from the California
	Institute of Technology, Pasadena, CA, USA, in
	1998. He served as a Faculty Member with
	the University of Minnesota, Minneapolis, MN,
	USA, then with the Texas A\&M University
	at Qatar, Doha, Qatar, before joining King
	Abdullah University of Science and Technology,
	Thuwal, Saudi Arabia, as a Professor of Electrical
	Engineering in 2009. His current research interests
	include the modeling, design, and performance analysis of wireless
	communication systems.
\end{IEEEbiographynophoto}

\end{document}

%% file: notation.tex
\def\nb0{{\mathbf{0}}}
\def\nb1{{\mathbf{1}}}

